\documentclass[%
 reprint,
 amsmath,amssymb,
 aps,
]{revtex4-2}

\usepackage{graphicx}
\usepackage{dcolumn}
\usepackage{bm}

\usepackage{physics}

\newcommand{\eu}{EuB$_6$}

\usepackage[x11names]{xcolor}

\begin{document}

\preprint{APS/123-QED}

\title{Does a low-carrier density ferromagnet hold the key to understanding high temperature superconductors?}

\author{Gabrielle Beaudin}
\affiliation{D\'epartement de Physique, Universit\'e de Montr\'eal, Montr\'eal, Canada}
\altaffiliation{Regroupement Qu\'eb\'ecois sur les Mat\'eriaux de Pointe (RQMP)}

\author{Alexandre D\'esilets-Benoit}
\affiliation{D\'epartement de Physique, Universit\'e de Montr\'eal, Montr\'eal, Canada}
\altaffiliation{Regroupement Qu\'eb\'ecois sur les Mat\'eriaux de Pointe (RQMP)}

\author{Andrea Daniele Bianchi}
\email{andrea.bianchi@umontreal.ca}
\affiliation{D\'epartement de Physique, Universit\'e de Montr\'eal, Montr\'eal, Canada}
\altaffiliation{Regroupement Qu\'eb\'ecois sur les Mat\'eriaux de Pointe (RQMP)}

\author{Robert Arnold}
\affiliation{School of Metallurgy and Materials, University of Birmingham, Elm Rd, Birmingham, B15 2TT, United Kingdom}

\author{Stavros Samothrakitis}
\affiliation{Laboratory for Neutron Scattering and Imaging, Paul Scherrer Institut, Forschungsstrasse 111, Villigen, CH-5232, Switzerland}

\author{Kilian D. Stenning}
\affiliation{Blackett Laboratory, Imperial College London, Prince Consort Road, London, SW7 2AZ,  United Kingdom}

\author{Mark Laver}
\affiliation{Department of Physics,  University of Warwick, Coventry, CV4 7AL,  United Kingdom}

\author{Simon Gerber}
\affiliation{Laboratory for X-ray Nanoscience and Technologies, Paul Scherrer Institut, Forschungsstrasse 111, Villigen, CH-5232, Switzerland}

\author{Nikola Anna Galvan Leos}
\affiliation{Laboratory for Neutron Scattering and Imaging, Paul Scherrer Institut, Forschungsstrasse 111, Villigen, CH-5232, Switzerland}

\author{Jorge L. Gavilano}
\affiliation{Laboratory for Neutron Scattering and Imaging, Paul Scherrer Institut, Forschungsstrasse 111, Villigen, CH-5232, Switzerland}

\author{Michel Kenzelmann}
\affiliation{Laboratory for Neutron Scattering and Imaging, Paul Scherrer Institut, Forschungsstrasse 111, Villigen, CH-5232, Switzerland}

\author{Michael Nicklas}
\affiliation{Physics of Quantum Materials, Max Planck Institute for Chemical Physics of Solids, N\"othnitzer Strasse 40, Dresden, 01187, Germany}

\author{Robert Cubitt}
\affiliation{Large Scale Structures Group,  Institut Laue Langevin, 71 avenue des Martyrs, Grenoble, 38042,  France}

\author{Charles Dewhurst}
\affiliation{Large Scale Structures Group,  Institut Laue Langevin, 71 avenue des Martyrs, Grenoble, 38042,  France}

\date{\today}

\begin{abstract}
We conducted a small-angle neutron scattering experiments (SANS) on the ferromagnetic semi-metal \eu, where we observed  direct evidence for the presence of magnetic polarons.
We carried out SANS experiments over a large range of scattering  vectors $\abs{\va{q}}$ from 0.006 to 0.140~\AA$^{-1}$ and  2 to 60~K. Just above $T_{\mathrm{C}}$ our experiments show magnetic scattering intensity, which has a Lorentzian dependence on the wave vector, which is characteristic for the presence of magnetic polarons.  Below $T_{\mathrm{C}}$ the polarons merge and most of the observed intensity is due to scattering from domain walls.
We were able to extract a correlation length $\xi$  which ranges from $100$ to $300$~\AA\ for the size of the magnetic polarons. This size is much larger than one would expect for magnetic fluctuations, demonstrating the  influence the magnetic polarons  on the phase transition. 
\end{abstract}

%

\maketitle


While we have a firm understanding of the physics of a high density electron gas, we have only begun to understand the physics at low carrier densities~\cite{ceperley_ground_1980,abrahams_colloquium_2001}. This is particularly true for the case, where the charge carriers are added to a background of localized magnetic moments. Contrary to intuition, the charge carriers will interact strongly, with the localized moments and form so-called magnetic polarons. 
Magnetic polarons can exist below $T_\mathrm{N}$ in antiferromagnets and in the paramagnetic phase of ferromagnets, as long as the carrier density is lower than the size of a magnetically correlated volume~\cite{majumdar_dependence_1998,Wegener2002,Calderon2004}. However, until now, they have not been directly observed. Their presence is expected in 
high temperature superconductors for example, where the parent compound is a strongly correlated antiferromagnetic Mott insulator, which then is typically doped with holes. These holes are mobile impurities which are moving in a background of strongly fluctuating magnetic moments. Due to strong interactions, they become dressed with a magnetic cloud, the magnetic polaron~\cite{emery_phase_1990,ashida_many-body_2018,grusdt_parton_2018}.
Furthermore, the distribution of these holes does not remain uniform, but they attract each other, leading to an electronic phase separation~\cite{emery_phase_1990}.  
This competition lies at the heart of a doping-dependent transition from an anomalous metal to a conventional Fermi liquid, and was recently observed in a quantum simulator of the Fermi-Hubbard model~\cite{koepsell_microscopic_2021}.
It is this strong interaction between the motion of the holes and antiferromagnetism that is believed to be at the heart of understanding the superconductivity in the cuprates~\cite{grusdt_parton_2018}. 

Magnetic polarons were  first proposed in the late sixties to describe the colossal magnetoresistive (CMR) effects in the Europium chalcogenides. Here, the magnetic polarons are thought to be formed by the spin polarization of the mobile charge carriers by  localized $4f$ moments. This leads to the formation of magnetic ``bubbles'' and the localization of charge carriers within them~\cite{VonMolnar1967,Kasuya1970,mauger_bound_1984,majumdar_dependence_1998,Wegener2002}. 
It is this CMR effect that could lead to new spintronic transistors~\cite{Prinz1998}. 

Here we present a small-angle neutrons scattering (SANS) study in ferromagnetic \eu, which we will argue is the ideal model system for studying magnetic polarons. Our results give for the first time  direct evidence  for magnetic polarons, and show their out-sized enhancement of the ferromagnetic spin fluctuations driving the electronic phase transition.

EuB$_6$ has a simple cubic crystal structure $(Pm\bar{3}m)$ but displays a complex interplay between the electronic and magnetic degrees of freedom due to its low carrier density~\cite{Wegener2002,Pereira2004,Calderon2004}. The insulator-to-semi-metal transition in EuB$_6$ is concomitant with a ferromagnetic phase transition~\cite{Henggeler1998}. A number of experiments have given indirect evidence for the presence of magnetic polarons in \eu~\cite{Nyhus1997,Sullow2000,Brooks2004,Das2012,Manna2014}.
A scanning tunnelling microscopy (STM) study showed that EuB$_6$   becomes electronically inhomogeneous for temperatures above  $T_\mathrm{C}$. Here, at  20~K the size of the inhomogeneities are of the order of 3 to 4~nm~\cite{Pohlit_STM_2018}. At the same time, measurements with a micro Hall-bar  pointed to  magnetic inhomogeneities at these temperatures~\cite{Pohlit_STM_2018}, which are pinned to defects at the surface~\cite{rosler_visualization_2020-1}. This is the very electronic phase separation expected for magnetic polarons.

Electronic inhomogeneity in \eu\ was also observed in an angle resolved magnetoresistance experiment~\cite{beaudin_possible_2022}. This concurrence of magnetic polarons and electronic inhomogeneity, as seen here in \eu\,  also manifests itself in the high-$T_c$'s. Here, a quantum nematic was first theoretically predicted for the doped two-dimensional Mott insulator~\cite{kivelson_electronic_1998}, and was later observed~\cite{cheong_incommensurate_1991,tranquada_coexistence_1997,ando_electrical_2002,hinkov_electronic_2008,ramshaw2017highTc}. In the high-$T_\mathrm{C}$'s the relation between nematic order and superconductivity, and its relation to  a close-by structural instability are hotly debated. Such a coupling of the quantum nematic to the lattice is absent in \eu~\cite{Booth2001}, which makes EuB$_6$ a clean experimental platform to study magnetic polarons.

EuB$_6$ is a magnetic semiconductor, which exhibits two phase transitions~\cite{Degiorgi1997}. Upon cooling from an insulating state at high temperatures, it first becomes a semi-metal, indicated by a drop in resistivity at $T_M$ of 14.5~K (see Fig.~3 of the Supplemental Material). At the Curie temperature $T_{\mathrm{C}}$ of 11.8~K, it orders ferromagnetically~\cite{Henggeler1998}.
EuB$_6$ displays CMR behaviour near $T_{\mathrm{C}}$~\cite{Sullow2000}. Also, \eu\ has a very low  carrier density~\cite{Wigger2004} of $\approx 10^{25}$~m$^{-3}$ at 20~K, which coexists with localized pure spin $4f$ Eu moments $(S=7/2)$. This puts EuB$_6$ into the regime where magnetic polarons are expected to strongly affect the electrical conductivity~\cite{majumdar_dependence_1998,Wegener2002,Calderon2004}. This scenario, is supported by a number of experiments~\cite{Brooks2004,Zhang2009,Das2012,Manna2014}. 

\begin{figure}[h!]
\centering
\includegraphics[width=\linewidth]{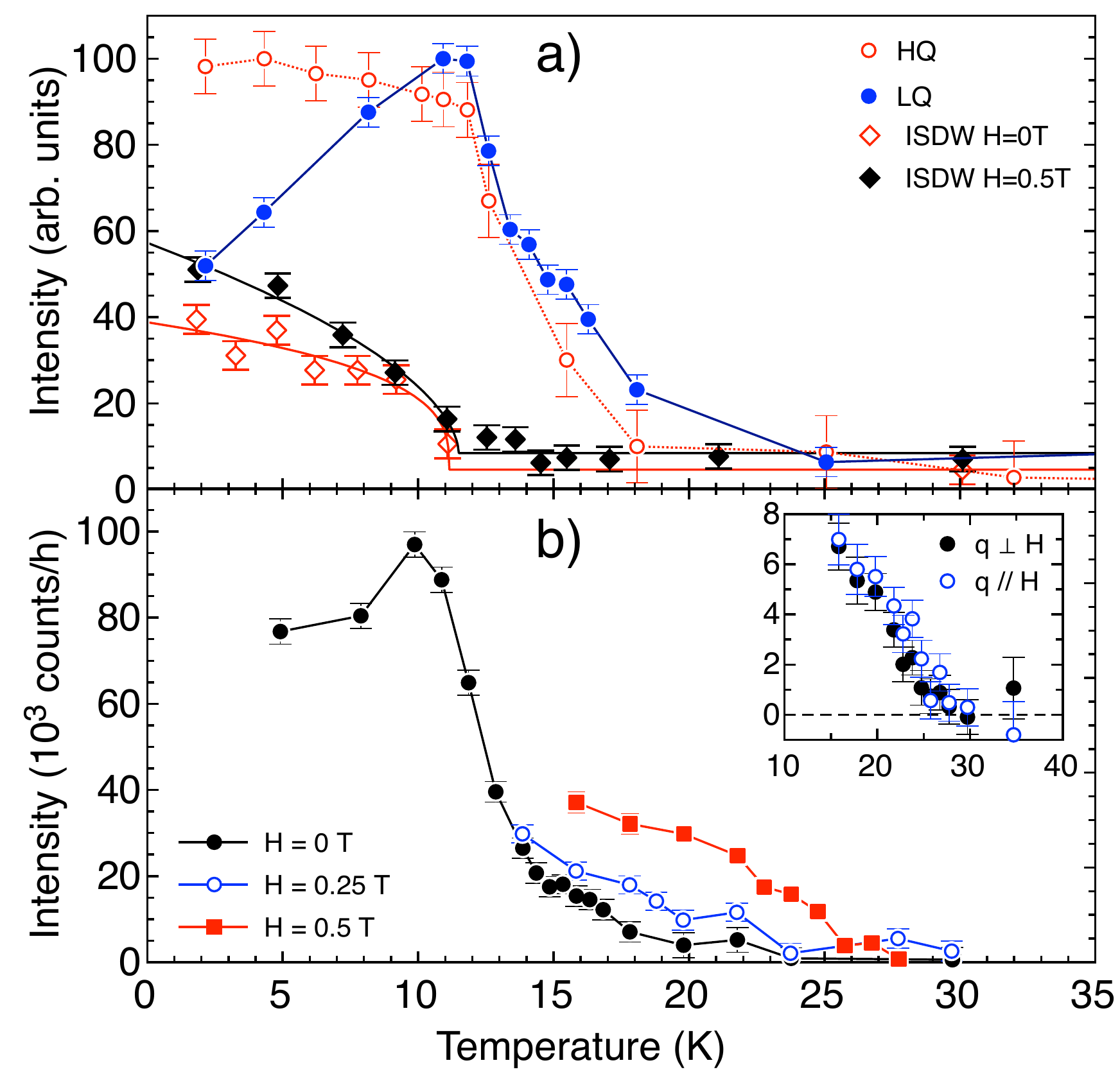}
\caption{
\textbf{a)} The solid black diamonds (empty red diamond) shows $I\qty(\abs{\va{q}})$ of the incommensurate magnetic peak at 0~T (0.1~T). The solid line is a fit to a power law. The solid blue circles are the data from the high $q$-regime (HQ) in ZF averaged over the $q$-range from 0.050 to 0.140~\AA$^{-1}$. The open red circles are the average over the low $q$ regime (LQ) in zero field from 0.006 to 0.025~\AA$^{-1}$. \textbf{b)} The medium $q$-regime is from 0.020 to $0.055$~\AA$^{-1}$(MQ). The full black circles show the ZF data, the open blue circles are taken in 250~mT and the full red squares in 500~mT. 
The inset shows the anisotropy of the SANS signal when $H$ is applied perpendicular to the neutron beam. 
}\label{fig:intensityvsT}
\end{figure}

 In our SANS experiment (see Fig.~1 of the Supplemental Material for a schematic), we probed the magnetic response of \eu\ for three different ranges of scattering vectors $q$. 
 We are able to distinguish three different sources of scattering in our SANS experiments. Firstly, diffuse scattering which grows in size with decreasing temperature, which is originating from magnetic polarons. 
 Secondly, we observe an incommensurate magnetic peak which appears below the Curie temperature $T_{\mathrm{C}}$. Thirdly, below $T_\mathrm{C}$ we observe a second diffuse scattering signal from ferromagnetic domain walls. An overview 
 is presented in Fig.~\ref{fig:intensityvsT}.

We start our discussion with the incommensurate magnetic peak which appears   below $T_{\mathrm{C}}$  as a spot on the detector. At 1.5~K, its position   corresponds to a $\abs{\va{q}}$ of $(18.4\pm0.2) \times 10^{-3}$~\AA$^{-1} $. This  $\abs{\va{q}}$-value  is at wavelengths above those required for double scattering and thus the peak is due to an incommensurate spin density wave (ISDW). An  ISDW is expected for \eu, as a group theoretical analysis using ISOTROPY~\cite{campbell_isodisplace_2006} for the crystallographic space group $Pm\overline{3}m$ of \eu\ with a $(100)$ propagation vector~\cite{Henggeler1998} yields  magnetic space groups with a  symmetry below that of the crystal~\cite{Sullow2000,beaudin_possible_2022}. The period of this ISDW ranges from 680~\AA\ to 726~\AA\ and its intensity corresponds to a magnetic moment of 2.5~$\mu_\mathrm{B}$/Eu at 2~K. An ISDW is characteristic for strongly correlated systems, as diverse as Ca-doped Sr$_2$RuO$_4$~\cite{SDW_SrRuO4}, a ruthenate, the La$_2$CuO$_4$ family~\cite{SDW_La2CuO4} of high-$T_c$'s, as well as BiFeO$_4$~\cite{Ramazanoglu_BiFeO3}, a multiferroic. What is common to all these systems is that their main order parameter is always found in strong competition with  other competing order parameters~\cite{SDW_La2CuO4}. This suggests a similar scenario for EuB$_6$, where ferromagnetic order competes with an electronic phase transition. We  then proceeded to measure the temperature dependence of the intensity of this peak, which is shown in Fig.~\ref{fig:intensityvsT}a. As function of the temperature, the peak appears at  $T_\mathrm{C}$. The solid line is a fit to a power law, resulting in a $T_\mathrm{C}$ of $11\pm$ $\beta=0.6\pm0.1$ for 0~T and a $\beta=0.3\pm0.1$ at 0.5~T. 

To study the temperature dependence of the diffuse magnetic signal, we integrated the contributions from a large area of the detector (see Fig.~2 of the Supplemental Material). For large momentum transfers, the $\abs{\va{q}}$-range is sensitive to magnetic fluctuations, which are enhanced by the magnetic polarons. Indeed, as shown in Fig.~\ref{fig:intensityvsT}a, we see an increase in the integrated signal with $q$ ranging from 0.050 to 0.140~\AA$^{-1}$ at 18~K, and a small bump at $T_M$
(see Fig.~3 of the Supplemental Material). Below $T_{\mathrm{C}}$, the magnetic fluctuations are reduced since the number of polarons decreases. Also, below $T_{\mathrm{C}}$, most of the intensity is due to scattering from  ferromagnetic domain boundaries which is best probed through low scattering vectors. This is why the signal remains constant below $T_{\mathrm{C}}$ for large momentum transfers (Fig.~\ref{fig:intensityvsT}a, HQ).

\begin{figure}[h!]
\centering
\includegraphics[width=\linewidth]{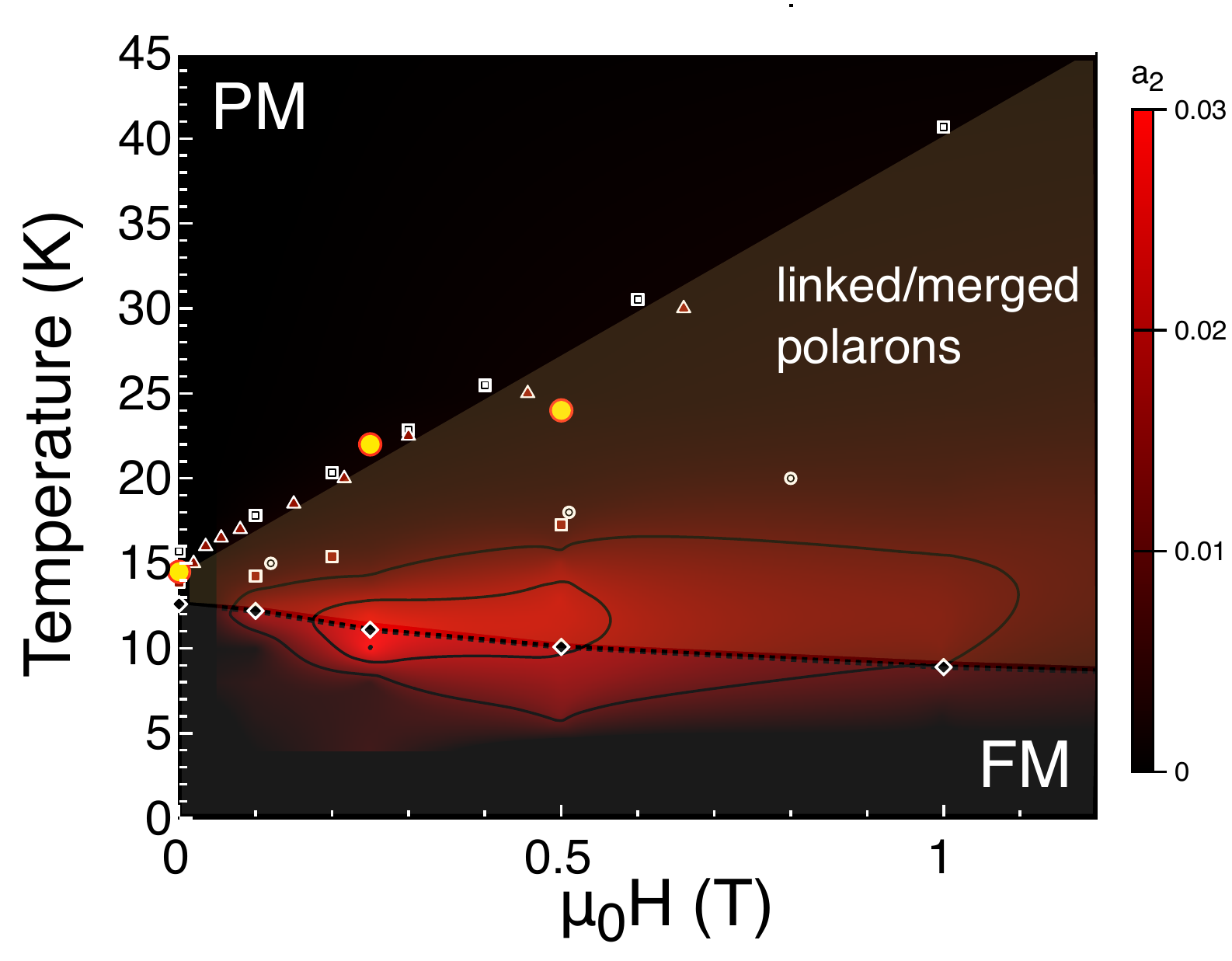}
\caption{
The solid yellow circles show the temperature at which the averaged intensity in the medium $q$-range starts to increase due to the formation of magnetic polarons. Their position agrees well with the region delineated by the the open black squares  from the Hall effect~\cite{Zhang2009}, as well as  the position of the peak in weak non-linear resistance measurements (red triangles)~\cite{Das2012}. The region of the $H-T$-diagram where  symmetry breaking observed in AMRO  is shown as a color plot~\cite{beaudin_possible_2022}. This region matches with the anomaly in magnetostriction~\cite{Manna2014}, and the position of $T_M$ identified in $\chi"$ measurements~\cite{beaudin_possible_2022}. The phase line $T_\mathrm{C}$ separating the ferromagnetic region was determined from $C_p$ and is shown as  solid black diamonds.
}\label{fig:phasediagram}
\end{figure}

We also tracked the onset of the signal from the magnetic polarons with magnetic fields. The data taken on the instrument D11 at the Institute Laue Langevin and averaged over the medium $q$-regime (0.020 to $0.055$~\AA$^{-1}$) is shown in Fig.~\ref{fig:intensityvsT}b, where we can see a shift in the uptake of the intensity towards higher temperatures: the bump marking $T_M$ is shifted from 16 to 22~K for a field of 0.25~T, and to 25~K for 0.5~T. This shift in temperature matches the boundary of the phase diagram of magnetic polarons extracted from noise spectroscopy measurements~\cite{Das2012}, and Hall effect~\cite{Zhang2009}, which confirms that the signal is due to magnetic polarons. The phase diagram is shown in Fig.~\ref{fig:phasediagram}. To complete the phase diagram, we used specific heat $C_p$ measurements in magnetic fields. As the peak in $C_p$  is strongly deformed by the application of a magnetic field, we used a fit to a mean field model for a ferromagnet to extract $T_\mathrm{C}$~\cite{Rodriguez_2005} (see Fig.~3 of the Supplemental Material). Critical exponents for the phase transition can be extracted from magnetic susceptibility measurements~\cite{Sullow2000, Sivananda_magneto_2018}. Our analysis yields the   exponents $\gamma$ of $1.03\pm 0.02$ and $\beta=0.49 \pm 0.02$ (see Fig.~4 of the Supplemental Material). These values are close to those of expected for a mean field model ($\gamma=1$ and $\beta=0.5$). This justifies the use of a mean field model to fit our $C_p$ data to extract $T_\mathrm{C}\qty(H)$, which is only weakly suppressed by the magnetic field (Fig.~\ref{fig:phasediagram}). 
We would also like to point out that there is no $\sin^2 \theta$ dependence visible in the SANS signal in the field-applied data (inset of Fig.~\ref{fig:intensityvsT}b), showing that the signal we see in EuB$_6$ is different from that expected in the case of random anisotropic ferromagnets~\cite{Michels2008}, or from superparamagnetic nanoparticles~\cite{Cywinski_superparamagnetic_1977}.  This suggests that the magnetic bubbles which we observe in \eu\ with SANS are not due to local static inhomogeneities, but they are rather dynamic fluctuations, as expected for magnetic polarons.  

\begin{figure}[h!]
	\centering
	\includegraphics[width=\linewidth]{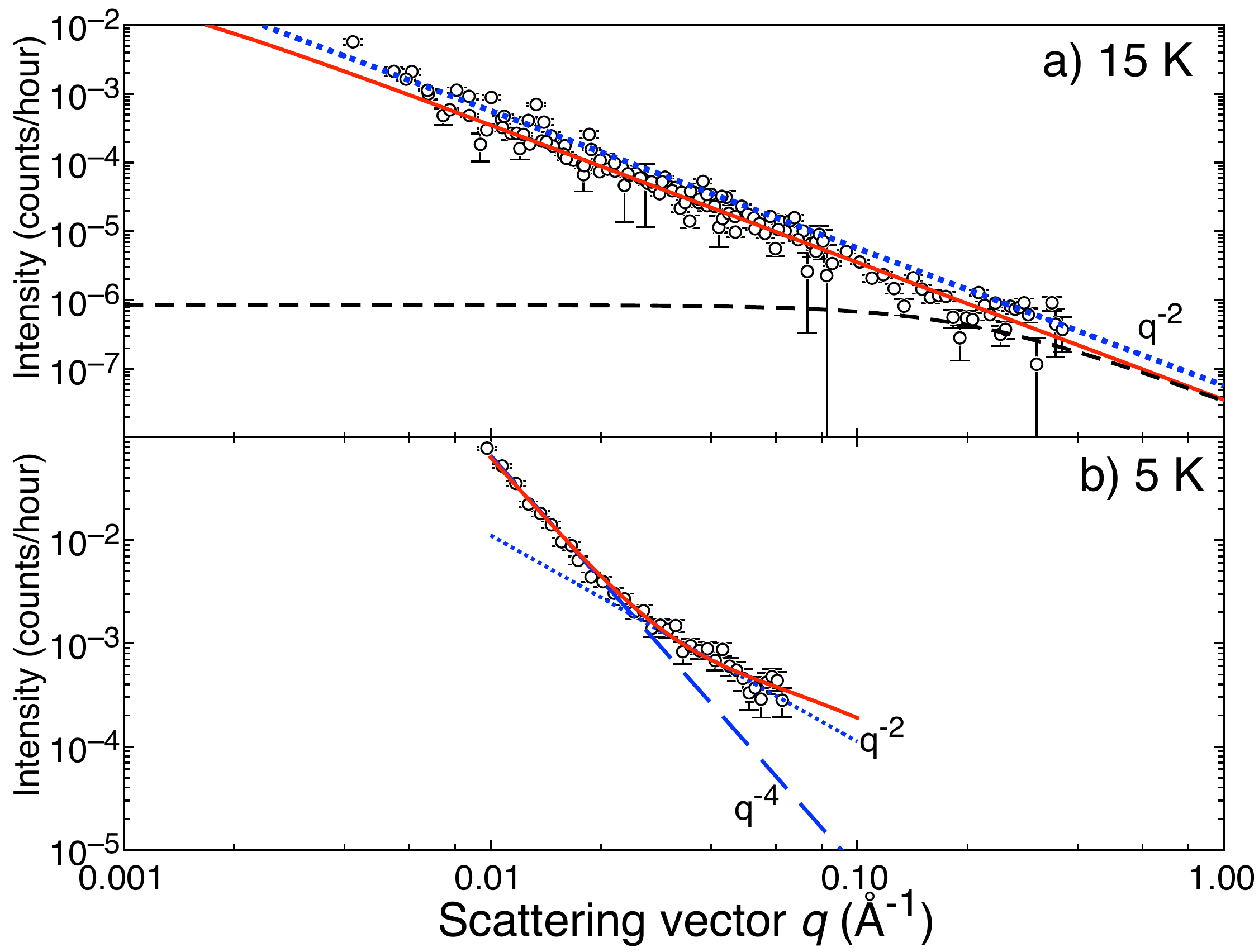}
	\caption{
	\textbf{a)} The upper panel shows $I\qty(\abs{\va{q}})$ data at 15~K. The data is described by  a Lorentzian function (red line). The dashed black line is the  contribution from 3D Heisenberg ferromagnetic fluctuations. \textbf{b)} At 5~K, below $T_{\mathrm{C}}$, the red line is a fit to a sum of a $q^{-4}$ and a Lorentzian, the dotted line is a guide to the eye showing the change in slope from $q^{-2}$~(small dots) to $q^{-4}$~(dash) at $q=0.024$~\AA$^{-1}$.}
	\label{fig:Ivsq}
\end{figure} 

Instead of integrating the intensity, we can inspect how the scattered intensity varies as a function of  $\abs{\va{q}}$. This allows us to distinguish the contributions of the magnetic polarons from those stemming from ferromagnetic domains. It also allows us to extract the correlation length for each contribution. For magnetic polarons,  we expect an intensity which varies as a Lorentzian~\cite{DeTeresa1997a}: 
\begin{equation}
 I_{\mathrm{pol}}(q) = \frac{A}{q^2+\kappa_p^2} \qquad,
\label{eq:Lorentzian}
\end{equation}
where $\kappa_p$ is the inverse correlation length $\xi$. 
This is also the $q$-dependence of magnetic fluctuations~\cite{Michels2008} close to a magnetic phase transition. This dichotomy can be resolved, as $\xi$ from magnetic fluctuations is only large for temperatures close to $T_M$, and it can be calculated using the critical exponents determined from the scaling plots.
In EuB$_6$ at 15~K and 0~T, the data over the entire $q$-range is described by a Lorentzian shown as the full line in Fig.~\ref{fig:Ivsq}a. 

\begin{figure}[h!]
	\centering
	\includegraphics[width=\linewidth]{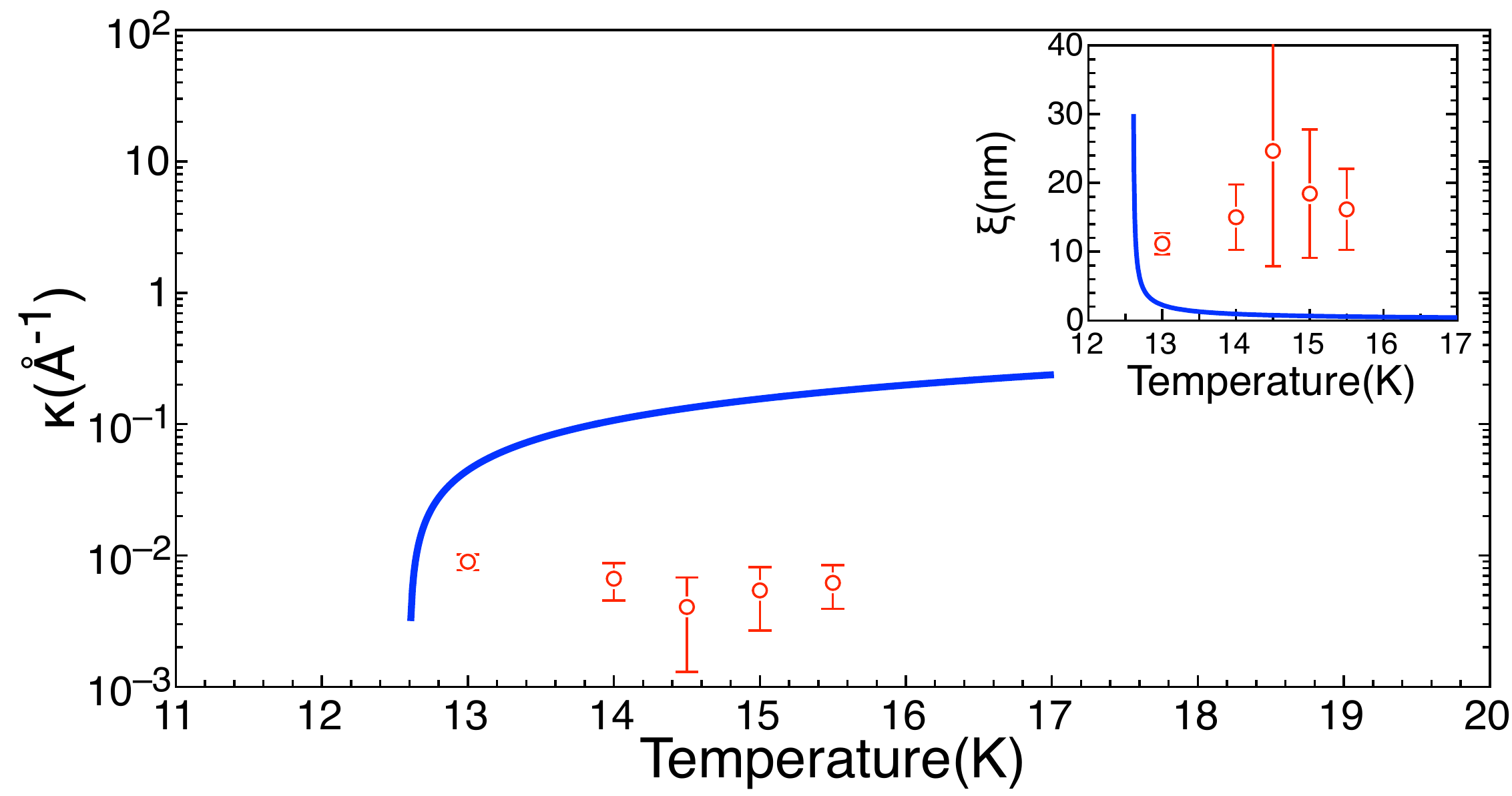}
	\caption{$\kappa =1/\xi$~vs.~$T$ of a 3D Heisenberg ferromagnet is shown as the solid blue (Dashed lines are the 10\% range around the estimated value).
	The open circles show $\kappa(T)$ obtained from fitting a Lorentzian. The inset shows  $\xi$~vs.~$T$. Again, the solid line is the calculation.}
	\label{fig:k1vsT}
\end{figure} 

$\xi(T)$  has a drastically different size and temperature dependence than what is expected for critical fluctuations. From the scaling of the magnetization, and the field dependence of $C_p$, we know that the  critical behavior of classical spin fluctuations in EuB$_6$ should be well described by a mean-field theory~\cite{Holm_criticalexpo_1993}. Thus, we can compute  $\xi(T)$, which is shown in the inset of Fig.~\ref{fig:k1vsT}. However, the experimental $\kappa(T)=1/\xi\qty(T)$ values  are much smaller, and have a different temperature dependence, than the estimate from the mean field model. This demonstrates that this signal is not due to classic ferromagnetic fluctuations, but that the magnetic fluctuations in \eu\ are strongly enhanced by magnetic polarons.
Moreover, the magnetic polarons, which drive the insulator to semi-metal transition, are also present at much higher temperatures than classical fluctuations. It can be seen in the inset of Fig.~\ref{fig:k1vsT}  that the size of the magnetic polarons starts to peak at $T_M$, while the classical magnetic fluctuations diverge at $T_\mathrm{C}$. The size of these compound objects is comparable to the size of the electronic inhomogeneities observed by STM~\cite{Pohlit_STM_2018}. 
We can also compare the size of the magnetic polarons to the volume occupied by a charge carrier. Using the charge carrier density $n$  at 20~K of $n\approx10^{25}$m$^{-3}$~\cite{Wigger2004}, we can estimate the volume per charge carrier to be $\sim 700$ unit cells or 50~nm$^3$, a size which again falls in the same range, as observed here by SANS, or by STM~\cite{Pohlit_STM_2018}. 

Below $T_\mathrm{C}$, the distribution of the size of the static regions of local spin ordering is expected to follow a Log-normal distribution~\cite{Wagner2007}.
This behavior has two possible origins: scattering from ferromagnetic domain walls~\cite{Hellman1999} or scattering from static ferromagnetic clusters~\cite{Marcano2007}. Here, the $q$-dependence is expected to follow a Lorentzian-squared function~\cite{Michels2008}:
\begin{equation}
I_{\mathrm{ferro}}(q)=\qty( \frac{B}{q^2+\kappa_F^2}) ^2 \qquad,
\label{eq:LorentzSquare}
\end{equation}
where, $\kappa_F$ is the inverse of the correlation length of the ferromagnetic domains~\cite{Hellman1999}. 
Below $T_\mathrm{C}$,  the characteristic length is set by the ferromagnetic domains. For very large domains the signal is proportional to $q^{-4}$.

Such a squared Lorentzian feature has been seen for example in  CeNi$_{1-x}$Cu$_x$, which is close to a ferromagnetic quantum critical point~\cite{Marcano2007}. Here, it was determined that the ferromagnetic order in CeNi$_{1-x}$Cu$_x$ arrives through the percolation of static ferromagnetic clusters. While this scenario bears similarity to what we observe in EuB$_6$, in CeNi$_{1-x}$Cu$_x$ the clusters are static and observed below the freezing point $T_f$ of a cluster glass transition, while in EuB$_6$ the magnetic polarons are dynamic and they are observed above $T_\mathrm{C}$ while showing a Lorentzian. The Lorentzian squared behavior we see in \eu\ below $T_\mathrm{C}$ is due to scattering by ferromagnetic domain walls~\cite{Hellman1999}. 
In \eu, the ferromagnetic domains below $T_\mathrm{C}$ lead  to a additional $q^{-4}$ contribution in addition to the $q^{-2}$ contribution from the dynamic magnetic polarons. This is shown in Fig.~\ref{fig:Ivsq}b. The size of  domains is larger than 100~\AA, which is about the largest size we can probe with our experiment.

In summary, we have showed evidence of magnetic polarons enhancing ferromagnetic fluctuations in EuB$_6$ using SANS for temperatures above the ferromagnetic transition. Below $T_\mathrm{C}$, the polarons merge and most of the observed intensity stems from scattering off domain walls
In addition, we determined that the polarons merge at the transition temperature $T_M=14.5$~K resulting in a drastic drop of the electrical resistivity. Applying a magnetic field shifts this percolation temperature to higher temperatures. The correlation length we see above the ordering temperature clearly indicates that the magnetic polarons strongly couple to the magnetic environment leading to strongly enhanced magnetic fluctuations. Our results put the existence of magnetic polarons in a ferromagnetic model system  on a firm footing  changing the status of these quasiparticles from the hypothetical to the real. Given the importance  of the electronic phase separation associated with magnetic polarons to spintronics and to high temperature superconductivity  we believe that our discovery of magnetic polarons which will lead to exciting new physics. 

\begin{acknowledgments}
The research at UdeM received support from the Natural Sciences and Engineering Research Council of Canada (Canada), Fonds Qu\'eb\'ecois de la Recherche sur la Nature et les Technologies (Qu\'ebec), and the Canada Research Chair Foundation. This work is based on experiments performed on the instruments SANS-I and SANS-II at the Swiss spallation neutron source SINQ, Paul Scherrer Institute, Villigen, Switzerland and on instruments D11 and D33 at Institut Laue-Langevin, Grenoble, France. The work at the MPI-CPFS was enabled through a DAAD grant.
\end{acknowledgments}

\end{document}